\documentclass{article}
\usepackage[utf8]{inputenc}

\usepackage{amsmath, amsthm, amssymb,color,fullpage,url,booktabs}  
\usepackage{graphicx}
\ifdefined\skiphyperref
\else
\ifdefined\blackandwhite
\usepackage[pdftex]{hyperref} 
\else
\usepackage[colorlinks,pdftex]{hyperref} 
\hypersetup{citecolor = blue, linkcolor = red!70!black}
\fi
\fi

\usepackage{cleveref}
\usepackage{tikz}
\usepackage{authblk}

\newcommand{\nc}{\newcommand}
\nc{\rnc}{\renewcommand}

\newcommand{\bra}[1]{\left\langle #1\right|}
\newcommand{\ket}[1]{\left|#1\right\rangle}

\def\be#1\ee{\begin{equation}#1\end{equation}}
\def\ba#1\ea{\begin{align}#1\end{align}}
\def\bas#1\eas{\begin{align}#1\end{align}}
\def\bpm#1\epm{\begin{pmatrix}#1\end{pmatrix}}
\nc{\non}{\nonumber}
\nc{\nn}{\nonumber}
\nc{\eq}[1]{(\ref{eq:#1})}
\nc{\eqs}[2]{(\ref{eq:#1}) and (\ref{eq:#2})}
\nc{\ra}{\rightarrow}
\nc{\ot}{\otimes}
\nc{\grad}{{\vec{\nabla}}}

\def\bea#1\eea{\begin{eqnarray}#1\end{eqnarray}}
\def\beas#1\eeas{\begin{eqnarray*}#1\end{eqnarray*}}

\makeatletter
\newtheorem*{rep@theorem}{\rep@title}
\newcommand{\newreptheorem}[2]{%
\newenvironment{rep#1}[1]{%
 \def\rep@title{#2 \ref{##1} (restatement)}%
 \begin{rep@theorem}}%
 {\end{rep@theorem}}}
\makeatother

\newreptheorem{theorem}{Theorem}
\newreptheorem{lemma}{Lemma}

\nc\eps{\epsilon}

\nc\cA{\mathcal{A}}
\nc\cB{\mathcal{B}}
\nc\cC{\mathcal{C}}
\nc\cD{\mathcal{D}}
\nc\cE{\mathcal{E}}
\nc\cF{\mathcal{F}}
\nc\cG{\mathcal{G}}
\nc\cH{\mathcal{H}}
\nc\cI{\mathcal{I}}
\nc\cJ{\mathcal{J}}
\nc\cK{\mathcal{K}}
\nc\cL{\mathcal{L}}
\nc\cM{\mathcal{M}}
\nc\cN{\mathcal{N}}
\nc\cO{\mathcal{O}}
\nc\cP{\mathcal{P}}
\nc\cQ{\mathcal{Q}}
\nc\cR{\mathcal{R}}
\nc\cS{\mathcal{S}}
\nc\cT{\mathcal{T}}
\nc\cU{\mathcal{U}}
\nc\cV{\mathcal{V}}
\nc\cW{\mathcal{W}}
\nc\cX{\mathcal{X}}
\nc\cY{\mathcal{Y}}
\nc\cZ{\mathcal{Z}}

\nc\bbC{\mathbb{C}}

\nc\bbF{\mathbb{F}}
\nc\bbM{\mathbb{M}}
\nc\bbN{\mathbb{N}}
\nc\bbR{\mathbb{R}}
\nc\bbZ{\mathbb{Z}}

\nc{\todo}[1]{\textcolor{red}{todo: #1}}
\nc{\Anote}[1]{\textcolor{red}{Aram note: #1}}

\newcommand{\vect}[1]{\boldsymbol{#1}}

\newcommand{\bv}{{\vect{b}}}
\newcommand{\vgamma}{\vect{\gamma}}
\newcommand{\vbeta}{\vect{\beta}}

\newcommand{\G}{\mathbb{G}}
\newcommand{\pr}{\mathbb{P}}

\newcommand{\ER}{Erd{\"o}s-R\'{e}nyi }

\def\begsub#1#2\endsub{\begin{subequations}\label{eq:#1}\begin{align}#2\end{align}\end{subequations}}
\nc\qand{\qquad\text{and}\qquad}
\nc\mnb[1]{\medskip\noindent{\bf #1}}

\nc{\pder}[2]{\frac{\partial {#1}}{\partial {#2}}}
\nc{\p}{\partial}
\title{The Quantum Approximate Optimization Algorithm Needs to See the Whole Graph: A Typical Case}
\author[1]{Edward Farhi}
\author[2]{David Gamarnik}
\author[ ]{Sam Gutmann}
\affil[1]{\small Google Inc., Venice CA 90291 and
Center for Theoretical Physics, MIT, Cambridge MA, 02139}
\affil[2]{\small Operations Research Center and Sloan School of Management MIT, Cambridge MA, 02140}

\begin{document}

\maketitle
\begin{abstract}
The Quantum Approximate Optimization Algorithm can naturally be applied to combinatorial search problems on graphs. The quantum circuit has p applications of a unitary operator that respects the locality of the graph. On a graph with bounded degree, with p small enough, measurements of distant qubits in the state output by the QAOA  give uncorrelated results. We focus on finding big independent sets in  random graphs with dn/2 edges keeping d fixed and n large. Using the Overlap Gap Property of almost optimal independent sets in random graphs, and the locality of the QAOA, we are able to show that if p is less than a d-dependent constant times log n, the QAOA cannot do better than finding an independent set of size .854 times the optimal for d large.  Because the logarithm is slowly growing, even at one million qubits we can only show that the algorithm is blocked if p is in single digits.  At higher p the algorithm ``sees" the whole graph and we have no indication that performance is limited.
\end{abstract}
  \section {Introduction\label{sec:Intro}}

The Quantum Approximation Optimization Algorithm \cite{FGG1,FGG2} is designed to
find approximate solutions to combinatorial search problems, and here
we consider its application to finding large independent sets in
random graphs. The graphs have $n$ vertices and $dn\over 2$ edges chosen uniformly at random, with $d$, the average degree of each graph, fixed.  The quantum algorithm
consists of an alternation of $2p$ unitaries, half of which are
single-qubit unitaries and the other half only interact qubits that
are connected by an edge in the graph.  On a bounded-degree graph,
with $p$ fixed or growing slowly with $n$, the QAOA does not ``see''
the whole graph.  This means that bits output by the QAOA are
uncorrelated at graph distances larger than $2p$. Looking at random graphs of average degree $d$, when $2p$ is less than a multiple of $\log n$ we will show that the power of the algorithm is limited. More precisely if $2p \le w\log n/\log(d/\ln 2) $ for any $w<1$  and $d$ big enough, the QAOA fails to produce an independent set larger than .854 times the optimal. (The ratio of logs is independent of the base of the log.) If $p$ is large enough that the algorithm ``covers" the whole graph we have no indication that the algorithm has limited power.

 In the first QAOA paper \cite{FGG1} it was shown that there exists a set of large Max-Cut instances on which the $p=2$ algorithm fails to achieve an approximation ratio of better than 0.756. These are bipartite 3-regular graphs with $o(n)$  squares. Although completely satisfiable, at this shallow depth, the QAOA can not detect if there are large odd length loops which would make it not completely satisfiable and hence the approximation ratio is provably less than $1$. Recently \cite{BKKT} looking at Max-Cut constructed a sequence of $d$-regular bipartite graphs for which the QAOA at depth $p<(1/3~ \mathrm{log}_2 n -4)d^{-1}$ fails to find an approximation ratio better than some $d$ dependent constant less than $1$. This result is similar to the one in this paper as it considers $p$ growing logarithmically with $n$. However theirs is a worst case result and ours is for typical instances.  Also crucially \cite{BKKT} require that the cost function have a $Z_2$ symmetry and that the initial state be an eigenstate of the $Z_2$ operator.  In our setup these conditions are not needed and for the problem we study they are not met.
 
 Our proof method uses the Overlap Gap Property (OGP) exhibited by the large independent sets of random graphs with bounded average degree established in~\cite{gamarnik2014limits}. Roughly speaking the OGP  says that for a given random graph, the intersection of any two large (\emph{i.e.} nearly optimal) independent sets is either big or small, that is, there is no middle ground.  The OGP was established to be an obstruction to a variety of classical algorithms, including local algorithms~\cite{gamarnik2014limits},\cite{gamarnik2017performance},\cite{chen2019suboptimality}, Markov Chain Monte Carlo (and related) methods, \cite{coja2017walksat},\cite{BAJag17},\cite{gamarnik2019overlap}, and Approximate Message Passing type algorithms~\cite{gamarnik2019overlapAMP}. The application of the OGP as a barrier to quantum algorithms is novel. It depends on the locality of the QAOA: the unitary operators in the algorithm only connect vertices which are connected in the input graph. As a result, because of the bounded average degree, when $p$ is a small multiple of $\log n$, changing a single edge of the graph affects only $o(n)$ qubits of the final state. We will use this to show that outputting a large independent set contradicts the OGP.
 
 It is worth remarking that the OGP is conjectured to \emph{not} hold for the low energy configurations of the Sherrington-Kirkpatrick model.  Assuming that the \emph{not} hold conjecture is true, Montanari recently  \cite{montanari2018optimization} constructed a polynomial time Approximate Message Passing  algorithm for finding a near ground state configuration.
 
 The QAOA has been applied to the Sherrington Kirkpatrick model at fixed $p$ in the infinite $n$ limit~\cite{farhi2019quantum}.  The associated graph is fully connected (each vertex has degree $n-1$) so the QAOA sees the whole graph at the lowest values of $p$. The techniques of this paper for bounded degree graphs cannot be applied to show any obstacle to the performance of the QAOA on this model.

   \section {Maximum Independent Set\label{sec:MIS}}
   
The computational problem we focus on is Maximum Independent Set or MIS.  Given a graph defined as a collection of vertices and edges, an independent set is a subset of the vertices with no graph edge between any two vertices in the set. It is easy to find a small independent set. Finding a big independent set is the challenge.  In fact finding the largest independent set in an arbitrary graph is an NP-hard problem.  But here we focus on random graphs and are interested in finding big independent sets but not necessarily the biggest. We define
$\alpha(G)$ as the independence ratio which is the size of the biggest independent set in $G$ divided by the number of vertices. Given a graph $G$, an algorithm will output an independent set and the quality of the algorithm can be measured by the size of the output divided by $n$ as compared to $\alpha(G)$.

We focus on sparse \ER graphs of $n$ nodes and $m$ edges where $m={d n\over2}$ with $d$ fixed, that is independent of $n$. In other words
$\G(n,{dn\over2})$ is a graph with ${d n\over2}$ edges chosen by picking this
number of edges uniformly at random  from the $n(n-1)/2$ possible
edges. The average degree of each graph is $d$.
The independence ratio $\alpha(\G(n,{d n\over2}))$ of this graph denoted by $\alpha_{n,d}$
is a random variable with the following known properties. First~\cite{BayatiGamarnikTetali} there exists an $\alpha_d$ such that
\begin{align} \label{eq:alpha1}
\alpha_{n,d}\to \alpha_d \text{ with probability $1$ as } n\to\infty.
\end{align}
Second, while the value of $\alpha_d$ for finite $d$ is unknown, the asymptotic value of $\alpha_d$ as $d$ increases 
is known~\cite{FriezeIndependentSet}:
\begin{align}\label{eq:alpha2}
\lim_{d\to\infty}{\alpha_d\over 2\ln d/d}=1.
\end{align}

We are interested in algorithms that take a graph as input and output a large independent set. The natural question that arises is if a polynomial time algorithm can produce independent sets close to  $\alpha_d$. 
A simple greedy algorithm achieves asymptotically  half of the optimal value, that is, it constructs an independent set of size $n(\ln d/d)(1+o_d(1))$ where $o_d(1)$ denotes a function of $d$  converging to $0$ as $d\to\infty$. Finding a polynomial time algorithm that provably goes beyond this would be a major achievement and it mirrors a similar problem in the context of dense \ER graphs, which has been open for more than four
decades~\cite{karp1976probabilistic}. Our interest is a quantum algorithm, the QAOA.  We will show that if $p$, the depth of the QAOA is less than $\epsilon$ log $n$ with $\epsilon$ a small constant, then the QAOA fails to go as far as $.854\alpha_d$ for $d$ large. For larger $p$ our arguments do not apply and we cannot say if the QAOA gets close to $\alpha_d$ with $p$ say growing as a large constant times log $n$. (Actually if we let $p$ grow fast enough with  $n$ the QAOA will find the optimum \cite{FGG1}.)

  \section{The Quantum Approximate Optimization Algorithm} \label{sec:QAOA}

We start by reviewing the QAOA with the graph problem Maximum Independent Set in mind. It is convenient here to work with bits that are $0,1$ valued. Given a classical cost function $C(\bv)$ defined on $n$-bit strings $\bv=(b_1,b_2,\ldots, b_n) \in \{0, 1\}^n$, the QAOA is a quantum algorithm that aims to find a string $\bv$ such that $C(\bv)$ is close to its absolute maximum.
The  graph-dependent cost function $C$ can be written as an operator that is diagonal in the computational basis, defined as
\begin{equation}
C\ket{\bv} = C(\bv)\ket{\bv}.
\end{equation}
We only consider ``local" cost functions, that is, those that only have interactions between qubits that are connected on the instance graph.
The problem dependent unitary operator depends on $C$ and a single parameter $\gamma$
\begin{align} \label{eq:UC}
U(C,\gamma) = e^{-i\gamma C}.
\end{align}
Note that $U(C,\gamma)$ conjugating a single qubit operator produces an operator that only involves that qubit and those connected to it on the graph.

The operator that induces transitions between strings uses
\begin{align}
B = \sum_{j=1}^n X_j,
\end{align}
where $X_j$ is the Pauli $X$ operator acting on qubit $j$, and the associated unitary operator  depends on a parameter $\beta$
\begin{align} \label{eq:UB}
U(B,\beta) = e^{-i\beta B} = \prod_{j=1}^n e^{-i\beta X_j}.
\end{align} Note that $U(B,\beta)$ conjugating a single qubit rotates that qubit and has no effect on other qubits.

We initialize the system of qubits in a product state such as 
\begin{align} \label{eq:psi0}
\ket{s} = \ket{0}^{\otimes n} 
\end{align}
or
\begin{equation}
\ket{s} = \ket{+}^{\otimes n} =  \frac{1}{\sqrt{2^n}} \sum_{\bv}\ket{\bv}.
\end{equation}
Using a product state for the initial state is the usual choice for the QAOA and is required for the arguments below.

We alternately apply $p$ layers of $U(C,\gamma)$ and $U(B,\beta)$.
Let $\vect\gamma = \gamma_1,\gamma_2,\ldots,\gamma_p$ and $\vect\beta = \beta_1,\beta_2,\ldots,\beta_p$. 
The QAOA circuit prepares the unitary operator
\begin{align} \label{eq:unitary}
U=U(B,\beta_p) U(C,\gamma_p) \cdots U(B,\beta_1) U(C, \gamma_1)
\end{align}
which acting on the initial state gives 
\begin{align} \label{eq:wavefunction}
\ket{\vect\gamma, \vect\beta} = U\ket{s}.
\end{align}
The associated QAOA objective function is
\begin{align}
\bra{\vgamma,\vbeta}C\ket{\vgamma,\vbeta}. \label{eq:QAOA_obj}
\end{align}
By repeatedly measuring the quantum state $\ket{\vgamma,\vbeta}$ in the computational basis, one will find a bit string $\bv$ such that $C(\bv)$ is near \eqref{eq:QAOA_obj} or better.
Different strategies have been developed to find optimal $(\vgamma,\vbeta)$ for any given instance~\cite{S1,S2}. But here we will show that under certain circumstances no set of parameters $(\vgamma,\vbeta)$ can achieve a certain level of success so we need not concern ourselves with optimal parameters. So from now on we denote the state produced by the QAOA as $\ket{\psi}$.

 \section{Locality Properties of the QAOA}

In this paper we are focusing on combinatorial search problem associated with random graphs of bounded  average degree $d$, so it it very unlikely that any vertex has degree much larger than $d$. This means that the cost function unitary \eq{UC} conjugating say a single qubit operator typically produces an operator acting on no more than roughly $d$ qubits.  The ``driver" unitary in the form of \eq{UB} introduces no spreading at all.  We also use for the initial state a product state which has no entanglement.  With this form of the QAOA we can establish some general locality properties of the quantum state produced by the quantum circuit. What follows is not restricted to a particular computational problem or to random graphs.

\subsection{Distant qubits}Consider an instance of some graph problem with its associated local cost function $C$. The first property has to do with bits that are far away from each other on the graph.  Define B($i,r$) as the set of vertices that are within a distance $r$ of the vertex $i$. Let Dist($i,2p$) be the complement of B($i,2p)$, that is, it is the set of vertices at least $2p$ away from $i$.  We are assuming 
than $2p$ is small enough so that Dist($i,2p$) is not empty. Let $O_i$ be an operator acting on qubit $i$ tensored with the identity acting on all other qubits.  Let $O_{\text{dist}}$ be an operator acting only on the qubits in Dist($i,2p$).  We now show that

\be
\bra{s} U^\dag O_i 
O_{\text{dist}} U \ket s =
\bra{s} U^\dag O_i U \ket s
\bra{s} U^\dag O_{\text{dist}} U \ket s
\label{eq:light-cone-1}\ee
as long as $\ket s$ is a product state and $U$ is of the form \eq{psi0} with a local cost function.

Proof of \eq{light-cone-1} .  Because the QAOA is local we see that $U^\dag O_i U$ only involves qubits in B($i,p$).  Because $O_{\text{dist}}$ only involves qubits that are $2p$ away from qubit $i$ we see that $U^\dag O_{\text{dist}} U$ can only involve qubits in the complement of B($i,p$).  Now

\be
\bra{s} U^\dag O_i 
O_{\text{dist}} U \ket s =
\bra{s} U^\dag O_i U  U^\dag O_{\text{dist}} U \ket s
\label{light-cone-2}\ee
and we will insert between $U$ and $U^\dag$  a complete set  with qubits in B($i,p$) and its complement.  Call $``\mathrm{near}"$ the set of bits in B($i,p$)  and those in its complement ``$\mathrm{far}$".  Now the initial state is a product state which we can write as $\ket{s}=\ket{s_\mathrm{near}}\ket{s_\mathrm{far}}$. Insert a complete set in the middle of the right hand side of \eqref{light-cone-2} and we get
\begin{align}
   =
  \sum_{\bf{v}_\mathrm{near}} \sum_{\bf{v}_\mathrm{far}} \bra{s_\mathrm{near}} \bra{s_\mathrm{far}} U^\dag O_i U  \ket{\bf{v}_\mathrm{near}} \ket{\bf{v}_\mathrm{far}} \bra{\bf{v}_\mathrm{near}} \bra{\bf{v}_\mathrm{far}} U^\dag O_\mathrm{dist} U \ket{s_\mathrm{near}} \ket{s_\mathrm{far}}
\end{align}
where the basis set $\ket{\bf{v}_\mathrm{near}}$ contains $\ket{s_\mathrm{near}}$ and the basis set $\ket{\bf{v}_\mathrm{far}}$ contains $\ket{s_\mathrm{far}}$. 
Now the $U^\dag O_i U$ term collapses the sum on 
$\bf{v}_\mathrm{far}$ and the $U^\dag O_{\text{dist}} U $ term collapses the sum on $\bf{v}_\mathrm{near}$ and we get
\be
 \bra{s_\mathrm{near}} \bra{s_\mathrm{far}} U^\dag O_i U \ket{s_\mathrm{near}} \ket{s_\mathrm{far}}\bra{s_\mathrm{near}} \bra{s_\mathrm{far}}U^\dag O_\mathrm{dist} U \ket{s_\mathrm{near}} \ket{s_\mathrm{far}}
\ee
which results in \eq{light-cone-1}.

In terms of the state $\ket{\psi}$ produced by the QAOA \eq{light-cone-1} says that
\begin{align} 
\bra{\psi} O_i O_\mathrm{dist}\ket{\psi}=\bra{\psi}O_i\ket{\psi} \bra{\psi} O_\mathrm{dist} \ket{\psi}.
\end{align}
Again $O_i$ is any operator acting on qubit $i$ and $O_\mathrm{dist}$ is any operator acting on qubits at least $2p$ away from $i$ and we see that the measurement outcomes of the two operators are independent. In particular  measurement in the state $\ket{\psi}$ of bit values at $i$ and in Dist are independent.
\subsection{Far from an edge}
For the next property imagine changing the cost function on a single edge. This locality property concerns the influence of this change on qubits that are far away from that edge.  Consider two instances of some computational problem which differ only by the presence or absence of a single edge. Or perhaps they differ because the single edge is weighted differently in the two instances. Call the associated cost functions $C$ and $C'$ which give rise to $U$ and $U'$ through \eq{UC}.  Let the edge in question be between vertices $i$ and $j$. Let Far$(ij,p)$ be the complement of $\mathrm{B}(i,p)\cup\mathrm{B}(j,p)$.  We assume that $p$ is small enough that Far$(ij,p)$ is not empty. Consider an operator $O_{\text{far}}$ that involves only qubits in Far$(ij,p)$. Now for the depth $p$ algorithm, $\ U^\dag O_{\text{far}} U$ does not involve the edge $ij$ so it is same with $U$ replaced by $U'$ that is

\be
\  U'^\dag O_{\text{far}} U'  = 
\  U^\dag O_{\text{far}} U .
\label{eq:light-cone-3}
\ee
What this means is that the influence of the edge in question is limited to qubits within a distance $p$ of the edge.  It also means that the probability of measuring a bit string in Far$(ij,p)$ is unaffected by the change in the edge $ij$.  Let $\ket{\psi}=U \ket{s}$ and $\ket{\psi'}=U'\ket{s}$ so these are the states produced with the unmodified and modified edge sets.  Consider $\textbf{b}_{\text{far}}$ which is the bit values of the set of bits in Far$(ij,p)$. Let $O_{\text{far}}= \ket{\textbf{b}_{\text{far}}}\bra{\textbf{b}_{\text{far}}}$ tensored with the identity on qubits in $\mathrm{B}(i,p)\cup\mathrm{ B}(j,p)$.  Now write $\ket{\bv}=\ket{\bv_\text{near}}\ket{\bv_\text{far}}$ and take the expectation of  \eq{light-cone-3} in the state $\ket{\bv}$. Keep $\textbf{b}_{\text{far}}$ fixed and sum on $\textbf{b}_{\text{near}}$ to get

\be \sum_{\bv_\mathrm{near}} 
\big|\langle{\bv_\text{near}}\big|\bra{\textbf{b}_\text{far}}\psi\rangle\big|^2= \sum_{\bv_\mathrm{near}}\big|\bra{\bv_\text{near}}\bra{\textbf{b}_\text{far}}\psi'\rangle\big|^2
\label{eq:light3}.
\ee
This means that the probability of measuring the bit string $\textbf{b}_{\text{far}}$ in Far$(ij,p)$ is unaffected by the edge change.
\subsection{Concentration of Hamming weight}

Again we are considering the QAOA with a local cost function associated with a graph $G$, a one local driver operator such as in \eq{UB}, and a product state for $\ket{s}$.  For fixed $p$ each vertex $i$ has a neighborhood B($i,2p)$. We take $p$ small enough that the maximum size of these neighborhoods is less than $n^A$ for some $A<1$. Run the QAOA to get the state $\ket{\psi}$ and measure in the computational basis to get a bit string. We now show that the Hamming weight of these measured bit strings concentrates in that, for sufficiently large $n$, each measurement produces the same Hamming weight with a variance that is $o(n)$.

Let the Hamming weight operator be 
\be\label{eq:ham}
W=\sum_{i} b_i.
\ee
The measurement variance of $W$ is 
\be
\bra{\psi}W^2\ket{\psi}-\bra{\psi}W\ket{\psi}^2
\ee
which breaks into $n^2$ terms

\be
\sum_{i}\sum_{j} \Big[\bra{\psi}b_ib_j\ket{\psi}-\bra{\psi}b_i\ket{\psi}\bra{\psi}b_j\ket{\psi}]~\Big].
\ee
Now for a fixed $i$ consider the sum on $j$.  If $j$ is more than $2p$ away from $i$ we can use \eq{light-cone-1} to see that this term is $0$.  So the only contributions can come from the $n^A$ nearby qubits and we bound the variance by $n^{(1+A)}$. The expected value of the Hamming weight scales with $n$ so the distribution concentrates.  

However we need a stronger result for our arguments. Let $G$ be a graph with $n$ vertices and as before  take $p$ small enough that the maximum size of B($i,2p$) is less than $n^A$ for some $A<1$ with high probability.  We are thinking of $n$ as large. For a given QAOA circuit we make the state $\ket{\psi}$ and measure the Hamming weight. Call the observed value $W_\text{obs}$. Then there exists $\gamma>0$ such that for every $\delta>0$ and for $n$ large enough
\begin{align}\label{eq:concentrate1}
\pr_G\Big[~\big| W_\text{obs}-\bra{\psi}W\ket{\psi}\big|\ge \delta n\Big]\le e^{-\delta n^\gamma},
\end{align}
where the graph is fixed and  the probability is over  measurements of $W$ in the state $\ket{\psi}$.  We are going to prove this in section 8 and also apply it to random graphs.

\section{QAOA applied to Maximum Independent Set}

We have discussed the QAOA in general and here we specify it for MIS.  For any graph we are looking for a big independent set, that is, a string with a large Hamming weight which corresponds to an independent set.  If we choose for the cost function the Hamming weight given by \eq{ham} we will easily discover strings with a big Hamming weight but they typically will not be independent sets of the input graph.  So we also consider the Independent Set cost function: 

\be\label{Cind} C_\text{IS}=\sum_{\langle i j \rangle}b_i b_j
\ee
where the sum is only over edges in the graph so $C_\text{IS}$ is local. We want a big Hamming  weight $W$ and $C_\text{IS}$ to be as small as possible. So for the objective cost function consider: 
 \be C_\text{obj} = W - C_\text{IS}.
 \label{eq:Cobj}
 \ee

When we run the QAOA our goal is to make $C_\text{obj}$ big.  But the cost function that appears in \eq{UC} need not be $C_\text{obj}$.  We can for starters take the cost function that appears in \eq{UC} to be $C_\text{IS}$ with the goal of making the quantum expectation of
$C_\text{obj}$ big. Regardless of what we take to drive this local QAOA, the strings that are output will not be independent sets of the associated graph. However these strings can be pruned to produce independent sets.

Suppose the quantum algorithm outputs a string with a positive value of $C_\text{obj}$.  By pruning we can produce an independent set of size at least this value. To see this consider the set of graph edges that exist between any of the vertices associated with $1$'s in the output string. Call the number of these edges $N_E$.  Now for each of these edges pick one of the two associated vertices at random and remove it from the string.  This reduces the Hamming weight by at most $N_E$ and reduces $C_\text{IS}$ by $N_E$ so $C_\text{obj}$ can not go down. This also shows that the maximum of $C_\text{obj}$ is the size of the largest independent set. We call the QAOA augmented by this pruning the QAOA+. Note that the pruning process is done randomly and respects the locality of the underlying graph.  When we run the QAOA+ at depth $p$ we mean that the QAOA is run at depth $p-1$ and the pruning is the last layer.

We now show that the shallowest depth version of the QAOA will produce a string with the objective function value being near $1.02 n/d$ for large $n$.  Here we pick for the starting state a rotation away from the all zeros state so that the initial Hamming weight is not zero.  Introduce a parameter $\theta$ and for $\ket{s}$ take

\be \ket{s} = U(B,\theta) \ket{0}
\ee
so the QAOA state is given by the three parameter unitary:

\be
\ket{\psi}=U(B,\beta)~U(C_\text{IS},\gamma)~U(B,\theta)\ket{0}
\ee
which we call the QAOA with $p=1.5$. Consider the simple case of $\gamma=0$ so the two rotations combine to be one rotation by $\theta+\beta$. This brings the state $\ket{0}$ to one where each bit $b_i$ has expected value $ \sin^2(\theta+\beta)$.  Now the expected value of $C_\text{obj}$ is $n [ \sin^2(\theta+\beta)-(d/2)\sin^4(\theta+\beta)]$ whose maximum is $1/2d$ at $\sin(\theta+\beta)=1/\sqrt{d}$.  Letting $\gamma$ vary can only improve this.  

For arbitrary $\theta$, $\gamma$ and $\beta$ we can evaluate the expectation of $C_\text{obj}$ in the state $\ket{\psi}$ by averaging over instances.  It is best to write \eqref{Cind} as
\be\label{Cind2} C_\text{IS}=\sum_{ij} J_{ij}b_i b_j
\ee
where each $J_{ij}$ is $1$ with probability $d/n$ and $0$ with probability $1-d/n$. Note that this is not exactly the same as the distribution we analyze in the rest of the paper which is a fixed number of edges $n d / 2$. But for the purposes of this calculation including edges with probability $d/n$ is more straightforward. In fact we can do the average over $J$ and get the expected value over graphs of the quantum expectation of the objective function
\begin{align}
    \mathrm{Ex}\big[ \bra{\psi} C_\text{obj} \ket{\psi}\big]
\end{align}
as $n$ times an explicit function of $d$ and the parameters $\theta$, $\gamma$ and $\beta$.  For each $d$ we can optimize numerically. For $d=3$ we get $.969 n / 3$. We see for large $d$ that \eqref{Cind2} approaches $1.02 n / d$ and the parameters $\theta$ and $\beta$ go down as $\sqrt{d}$. Pulling out a $1 / d$ and rescaling $\theta$ and $\beta$ we have a function that has a limit as $d$ goes to infinity.  The optimum of this function is $1.02$. 

What we have shown is that there is a form of the QAOA that we call the QAOA+ which at its lowest depth finds an independent set whose size is at least a constant times $n/d$. There are simple classical algorithms that can beat this. Our goal here was to show that the lowest depth QAOA+ can be analyzed on random instances and that it is a good starting point for understanding the QAOA at higher depth.

\section{ The Overlap Gap Property}\label{section:OGP}

We now describe the Overlap Gap Property which is the key to showing the limitation of the QAOA on random graphs. For random graphs, this property is satisfied by independent sets with size larger than a certain multiplicative factor $\eta^*\in (0,1)$
away from optimality. Specifically, this will be done for 
\begin{align}\label{eq:alpha^*}
\eta^*={1\over 2}+{1\over 2\sqrt{2}} = .853...  .
\end{align}
For each $\eta\in (0,1)$ and a graph $G$ on $n$ nodes let
\begin{align}
\mathcal{I}(\eta,G)=\{\sigma\in\mathcal{I}(
G): |\sigma|\ge n\eta\alpha_d\}
\end{align}
where $\mathcal{I}(G)$ is the set of all independent sets.
That is, $\mathcal{I}(\eta,G)$ is the set of independent sets in $G$ which are of size at least  $\eta$ times the asymptotic optimal. We call these large independent sets $\eta$-optimal.
 For any two $n$ node graphs $G_1$ and $G_2$, and for every $0<\tau\le \eta$, 
let $\mathcal{O}^{Ab}(\eta,G_1,G_2,\tau)$ denote the set of pairs $\sigma_1,\sigma_2$ such that $\sigma_j\in \mathcal{I}(\eta,G_j), j=1,2$ and
\begin{align}
|\sigma_1\cap\sigma_2|\ge n\tau\alpha_d,
\end{align}
that is, it is the set of pairs of $\eta$-optimal independent sets 
whose intersection size normalized by the asymptotic size of the largest independent set is above $\tau$.
Similarly, let
$\mathcal{O}^{Be}(\eta,G_1,G_2,\tau)$ denote the set of pairs $\sigma_1,\sigma_2$ such that $\sigma_j\in \mathcal{I}(\eta,G_j), j=1,2$ and
\begin{align}
|\sigma_1\cap\sigma_2|\le n\tau\alpha_d,
\end{align}
that is, it is the set of pairs of $\eta$-optimal independent sets 
whose intersection size normalized by the asymptotic size of the largest independent set is below $\tau$.

Let $G_0$  and  $G_m$ be chosen independently with distribution $\G(n,{d n\over2})$. We are going to introduce an interpolation between these two graphs. Let $(i^a_1,j^a_1),  \ldots, (i^a_m,j^a_m)$ with  $a=0,m$ be the corresponding
sets of $m$ edges of the two graphs. For every $t=0,1,2,\ldots,m$ consider an interpolating random graph $G_t$ 
with edges
\begin{align}
(i^0_1,j^0_1),  \ldots, (i^0_{m-t},j^0_{m-t}),(i^m_{m-t+1},j^m_{m-t+1}),\ldots, (i^m_m,j^m_m).
\end{align}  
$G_t$ uses the first $m-t$ edges from  $G_0$ and the remaining
$t$ edges from $G_m$.
For each fixed $t$, the graph $G_t$ is distributed as $\G(n,m)$, modulo the potential repetitions of edges. With high probability, the total number of edge repetitions is $O(1)$ with respect to $n$ so they can be ignored. 

\vspace{10pt}
\noindent\textbf{Theorem: Overlap Gap Property}
\textit{
For every $\eta>\eta^*$ there exists $0<\tau_1<\tau_2<\eta, d_0$ and $c>0$ such that for all $d>d_0$ 
\begin{align}
\mathrm{Prob}\Big[\exists \quad 0 \le t_1,t_2\le m \quad s.t. \quad
\mathcal{O}^{Ab }(\eta,G_{t_1},G_{t_2},\tau_1)\cap \mathcal{O}^{Be}(\eta,G_{t_1},G_{t_2},\tau_2)
\ne\emptyset\Big]\le \exp(-cn), \label{eq:OGP}
\end{align}
and
\begin{align}
    \mathrm{Prob}\Big[\mathcal{O}^{Ab}(\eta,G_{0},G_{m},\tau_1)\ne \emptyset\Big]\le \exp(-cn). \label{eq:near-orthogonal}
\end{align}
for all large enough $n$.}

\vspace{5pt}
The theorem makes two claims. First it says that across all pairs $G_{t_1},G_{t_2}$
of graphs in the interpolating sequence, every   $\eta$-optimal independent set $\sigma_1$ in $G_{t_1}$ and 
every   $\eta$-optimal independent set $\sigma_2$ in $G_{t_2}$
 have normalized intersection either
at most $\tau_1$ or at least $\tau_2$, except for an exponentially small probability. There is essentially no middle ground of pairs whose normalized intersection size is between $\tau_1$ and $\tau_2$. This is the Overlap Gap Property. The second says, 
for two independent random graphs sampled from $\G(n,{d n\over2})$ all corresponding pairs of large independent sets have normalized intersection at most $\tau_1$. For a proof and further discussion see \cite{gamarnik2014limits}.


\section{Overlap Gap Property is an obstruction to the QAOA+}
\subsection{Main Result}
\vspace{5pt}

Our main result is that the QAOA+, which is the QAOA augmented by pruning to produce independent sets,  applied to random graphs of average degree $d$ will fail to find an independent set close to optimal for $p$ less than a  constant times $\log n$. More precisely:

\vspace{10pt}
\noindent\textbf{Obstruction Theorem}
\vspace{5pt}

\textit{
For every $w < 1$ and $\eta >\eta^*$, there is a $\gamma > 0$ and a $d_0$ such that for $d>d_0$ we have:
If the QAOA+ is run on a $\G(n, {d n\over2})$ graph with 
\begin{align}
2p\le {w \log n\over \log(d/\ln{2})},
\end{align}
 then the probability that the algorithm outputs an independent set of size at least $\eta \alpha_d n$ is at most $e^{-n^\gamma}$ for all $n$ sufficiently large.}
\vspace{10pt}

We will prove this after stating some preliminary results. The first concerns the size of the neighborhoods of vertices in random graphs with $m={nd\over 2}$ edges.

\subsection{Preliminary Results}

\noindent\textbf{Neighborhood Size Theorem}
\vspace{5pt}

\textit{Fix $d>1$, and $w<1$. If
\begin{align}
2p\le {w \log n\over \log (d/\ln{2})},
\end{align}
then there exist $a>0$ and $A<1$ such that
\begin{align}
&\mathrm{Prob}\Big[\max_i\mathrm{B}(i,2p) \ge n^{A}\Big] \le e^{-n^a}  \\ and~~~
&\mathrm{Prob}\Big[\max_i\mathrm{B}(i,p) \ge n^{A/2}\Big] \le e^{-n^{a/2}}.
\end{align}
}

\noindent Here Prob is with respect to the graph distribution. What this says is that if $p$ is smaller than a certain constant times $\log n$, then in a random graph the  neighborhood ball of each vertex  contains a fraction of the vertices that vanishes as $n$ goes to infinity. We will prove this theorem in the next section. 

The next result is an expansion of the purely quantum result in section 4.2 to the QAOA+ which is the depth $p-1$ QAOA run with pruning so all outputs are independent sets.  The result \eq{light3} says that if a single edge in a graph is modified then the measurement probabilities of qubits far away are unaffected by the edge change. This is a statement about the QAOA but the QAOA outputs strings that need to be pruned back to make independent sets. The pruning process is random and respects the locality of the underlying graph. Let $\pr_G(\sigma)$ be the probability that the independent set $\sigma$ is output by the QAOA+ running on graph $G$.  Here the symbol $\pr_G$ means the graph $G$ is fixed and the randomness comes from the quantum measurement and the randomized pruning.  So we now state

\vspace{10pt}
\noindent\textbf{Far from an edge Lemma}

\vspace{5pt}

\textit{Consider an arbitrary graph $G$ with $G'$ obtained by adding a single edge $(i,j)$ to $G$. Let $\mathrm{Far}$ be the complement of $\mathrm{B}(i,p)\cup \mathrm{B}(j,p)$ where $p$ is the depth of the QAOA+.
Then for every }  $\sigma\in\{0,1\}^n$
\begin{align}
\sum_{\hat\sigma:\hat\sigma_k=\sigma_k, k\in \mathrm{Far}}\pr_{G}(\hat\sigma)=
\sum_{\hat\sigma:\hat\sigma_k=\sigma_k, k\in \mathrm{Far}}\pr_{G'}(\hat\sigma). \label{eq:equal-marginals}
\end{align}

\vspace{10pt}
\noindent Here we are assuming that $p$ is small enough that Far is not empty.
The proposition says that the total probability of independent sets which ``agree" with $\sigma$ on the node set Far, that is those nodes far from  the newly added
edge, remains the same after the  addition of the edge. The proposition is the direct implication of the local property \eq{light3} augmented by the fact that pruning is local.

Our next result is concerns the concentration of Hamming weight discussed in section 4.3 now extended to the QAOA+.

\vspace{10pt}

\noindent\textbf{Concentration Theorem}
\vspace{5pt}

\textit{1. Let $G$ be a graph with  $n$ vertices and ${d n\over2}$ edges. Suppose $p$ is chosen such that  $\max_{i }|\mathrm{B}_{G_n}(i,2p)|\le n^A$ for some $0 < A < 1 $.  Let $\sigma$ be the output of the QAOA+. 
Then there exists $\gamma_1>0$ such that for all $\delta>0$}
\begin{align}\label{c1a}
\pr_{G}\Big[~\big ||\sigma|-\mathbb{E}_{G}|\sigma|\big|\ge \delta n~\Big]\le e^{-\delta n^{\gamma_1}}
\end{align}
\textit{ for $n$ large enough. Here the graph is fixed and the randomness comes from the QAOA+.}

\textit{2.  Now let $G$ be a random $\G(n,{dn\over2})$ graph. Suppose for some $a>0$}, 
\begin{align}
&\mathrm{Prob}\Big[\max_i\mathrm{B}(i,2p) \ge n^{A}\Big] \le e^{-n^a} \  \\  
and~~~&\mathrm{Prob}\Big[\max_i\mathrm{B}(i,p) \ge n^{A/2}\Big] \le e^{-n^{a/2}}.
\end{align}
\textit{Then   there exists $\gamma_2>0$ such that for any $0<\delta<1$}, 
\begin{align}\label{c3a}
\mathrm{Prob}\Big[~\big| |\sigma|-\mathrm{Ex}|\sigma|\big|\ge \delta n~\Big]\le e^{-{\delta^2}n^{\gamma_2}}
\end{align}
\textit{for $n$ large enough. Here $\mathrm{Prob}$ and $\mathrm{Ex}$ refer to Probability and Expectation coming from the distribution of random graphs combined with the randomness coming from the QAOA+}.
\vspace{5pt}

This key concentration result will be proven in the next section. 
We now prove our main result.
\subsection{Proof of Obstruction Theorem}

Choose $w < 1$. Find  $A<1$  and $a$ as per the Neighborhood Size Theorem. 
Consider the interpolation $G_t, 0\le t\le m$ described in the previous section. Denoting by $\mathrm{B}_{G_t}(i,p)$
the neighborhood of node $i$ in graph $G_t$, we have by the union bound
\begin{align}\label{Bound-D}
\mathrm{Prob}\left[\max_t \max_{i}|\mathrm{B}_{G_t}(i,p)|\ge n^{A/2}\right]\le {d n\over2}e^{-n^{a/2}}\le e^{-n^{a'}},
\end{align}
where $a'$ is any constant smaller than $a\over2$ and $n$ is large enough. Let $D=\max_t\max_{i} |\mathrm{B}_{G_t}(i,p)|$.
On the sequence of graphs $G_t$ we construct a coupled sequence of independent sets $\sigma_t,  0\le t\le m$,
where for each fixed $t$, $\sigma_t$ is a single independent set coming from the distribution of the output of the QAOA+ on the random graph $G_t$. The set-theoretic difference $\Delta$ of
$\sigma_t$ and $\sigma_{t+1}$ will be seen to satisfy
\begin{align}
|\sigma_t\Delta \sigma_{t+1}|\le 4D. \label{eq:Bounded-sigma-difference}
\end{align}
By ``construct'' we do not mean an algorithmically efficient construction, rather we show that such a coupled sequence exists.
First produce a single sample $\sigma_0=(\sigma_{0,1},\ldots,\sigma_{0,n})$ by running the QAOA+ on $G_0$. Next, recall that $G_1$ is obtained from $G_0$
by deleting $G_0$'s last edge  $(i^0_m,j^0_m)$ and adding $G_m$'s last edge $(i^m_m,j^m_m)$. Consider the set of nodes 
\begin{align}
S=\mathrm{B}_{G_0}(i^0_m,
p)\cup \mathrm{B}_{G_0}(j^0_m,p)\cup \mathrm{B}_{G_1}(i^m_m,p)\cup \mathrm{B}_{G_1}(j^m_m,p).
\end{align}
Note  that $S$ has size at most $4D$. Find a sample $\sigma_1$ according to the conditional distribution 
\begin{align}\pr_{G_1}(\sigma_1 |\sigma_{1,i}=\sigma_{0,i}~
 i\notin S). 
\end{align}As a result the cardinality of $\sigma_0\Delta\sigma_1$ is at most $4D$. In the same fashion, we produce samples $\sigma_2,\ldots,\sigma_m$,
where each $\sigma_t$ is found by conditioning on $\sigma_{t-1}$ similarly. Again, $|\sigma_{t}\Delta\sigma_{t+1}| \le 4D$ for all $t=0,\ldots,m-1$.

Next, we claim that
$\sigma_1$ is distributed as $\pr_{G_1}$. Indeed for every $\sigma_1\in \{0,1\}^n$, its probability mass according to this sampling procedure is
\begin{align}
\sum_{\sigma_0: \sigma_{0i}=\sigma_{1i}, i\notin S} \pr_{G_0}(\sigma_0)\Bigg[ \frac{\pr_{G_1}(\sigma_1)}{\sum_{\sigma:\sigma_{i}=\sigma_{1i}, i\notin S} \pr_{G_1}(\sigma)}\Bigg]=\pr_{G_1}(\sigma_1)
\end{align}
by property (\ref{eq:equal-marginals}). Similarly, we have that $\sigma_t$ is distributed as 
$\pr_{G_t}(\sigma_t)$. We have shown that our desired coupled sequence of $m+1$ independent sets exists.

Note that 
$\mathrm{Ex}[|\sigma_t|]$
is independent of $t$, where as before the expectation is both with respect to the randomness of $G_t$ and the QAOA+. 
We claim that for every $\mu>0$ 
\begin{align}\label{eq:Bound-expectation}
\mathrm{Ex}[|\sigma_t|]\le (1+\mu)n\eta^*\alpha_d,
\end{align}
for all large enough $n$. Observe that by the Concentration Theorem it suffices to show (\ref{eq:Bound-expectation})
in order to prove our main result which is that the QAOA+ fails to produce independent sets that are bigger than $\eta^*$-optimal.

 Assume to the contrary, that  (\ref{eq:Bound-expectation}) is violated for infinitely many $n$.
For any such $n$, given $G_t$, $\sigma_t$ has the distribution $\pr_{G_t}$, and $G_t$ has the $\G(n,m)$ distribution.  Taking $\delta={\mu\over2}\eta^* \alpha_d$ in \eqref{c3a} and then using the union bound, we have
\begin{align}\label{eq:sigma_t-large}
\mathrm{Prob}\left[\min_t|\sigma_t|\ge (1+\mu/2)n\eta^*\alpha_d\right]\ge 1-e^{-n^{\gamma_3}},
\end{align}
for some $\gamma_3>0$. Let $\hat\eta=(1+\mu/2)\eta^*$. Find $0<\tau_1<\tau_2<\hat\eta$ from the Overlap Gap Theorem with respect to $\hat\eta$.
Then because $\tau_2 < \hat\eta$ we have
\begin{align}
\mathrm{Prob}\left[\min_t|\sigma_t|\ge n\tau_2\alpha_d\right]\ge 1-e^{-n^{\gamma_3}}.
\end{align}
By (\ref{eq:sigma_t-large}) and the second part of the Overlap Gap Theorem then 
\begin{align}
\mathrm{Prob}\big[~|\sigma_0\cap\sigma_m|\le n\tau_1\alpha_d~\big]\ge 1-e^{-n^{\gamma_3}}-e^{-cn}.
\end{align}
Now let us track the intersection $|\sigma_0 \cap \sigma_t|$ as $t$ goes from $0$ to $m$. For $t$ at $0$ this is bigger than $n \tau_2 \alpha_d$ but at $t=m$ it is less than $n \tau_1 \alpha_d$. However by the Overlap Gap Theorem we know there is (with high probability) no middle ground so there is some $T$ where $|\sigma_0 \cap \sigma_T|$
is big but $|\sigma_0 \cap \sigma_{T+1}|$ is small. As a general property of sets we have
\be \big| |\sigma_0 \cap \sigma_T| - |\sigma_0 \cap \sigma_{T+1}| \big| \le \big|\sigma_T\Delta\sigma_{T+1} \big|.
\ee
Using \eq{Bounded-sigma-difference} we get $n\alpha_d(\tau_2-\tau_1) \le 4D$ which is a contradiction for large enough $n$ because $D\le n^{A/2}$, $A/2<1/2< 1$, with high probability, see \eqref{Bound-D}.  This contradiction shows that the Obstruction Theorem is true.

\section{Proofs of Neighborhood Size and Concentration Results}\label{section:concentration}

\subsection{Neighborhood Size}

We begin by restating and then proving the Neighborhood Size Theorem. This result concerns random graphs and makes no reference to the quantum algorithm.
\vspace{5pt}

\noindent\textbf{Neighborhood Size Theorem}

\vspace{5pt}
\textit{Fix $d>1$, and $w<1$. If
\begin{align}
2p\le {w \log n\over \log (d/\ln{2})}
\end{align}
then there exist $a>0$ and $A<1$ such that
\begin{align}
&\mathrm{Prob}\Big[\max_i\mathrm{B}(i,2p) \ge n^{A}\Big] \le e^{-n^a} \ \\and~~~
&\mathrm{Prob}\Big[\max_i\mathrm{B}(i,p) \ge n^{A/2}\Big] \le e^{-n^{a/2}}.
\end{align}
}

To prove this we consider a branching process where each parent has Poisson($d$) children.
Let  $Z_k =$ the  size  of  the $k$th generation with  $Z_0 =1$.
Let
\begin{align}
\phi_k (t) = \mathrm{E} [e^{tZ_k}]
\label{overlap-eq38}
\end{align}
where the expectation E is with respect to the Poisson process.
We have
\begin{align}
 \phi_0(t) &= e^t\nonumber\\
\phi_1(t) &= e^{d(e^t-1)},
\label{overlap-eq39}
 \end{align}
the Poisson moment generation function, and generally
\begin{align}
\phi_{k+1}(t) = e^{d (\phi_k (t) -1)} .
\label{overlap-eq40}
 \end{align}
We first show 
\begin{align}
    \phi_k(({\ln{2}/ d})^k)\le  e \ \text{     for any} \ k \ge 0.
    \label{overlap-eq41}
\end{align}
Assume by induction on $j$, that
\begin{align}
    \phi_j(({\ln{2}/ d})^k)\le e^{(\ln{2}/ d)^{k-j}}
    \label{overlap-eq42}
\end{align}
which we show as follows.
This holds for $j=0$, and 
\begin{align}
    \phi_{j+1}(({\ln{2}/ d})^k)&= e^{d(\phi_j(({\ln{2} / d})^k) -1)}\nonumber\\
    &\le e^{d(e^{({\ln{2} / d})^{k-j}} -1)}
    \label{overlap-eq43}
\end{align}
by hypothesis.
Using $(e^{x\ln{2}} -1)\le x$ for $0\le x\le 1$, this is 
\begin{align}
\le e^{d (\ln{2})^{k-j-1} \over d^{k-j}} =  e^{(\ln{2} / d)^{k-(j+1)}}.
\label{overlap-eq44}
\end{align}

Now Markov's inequality says that for any $t$ and $u$
\begin{align}
    \mathrm{P} \left[Z_k\ge u\left({d/ \ln{2}}\right)^k\right]&\le 
    e^{-u\left(d/\ln{2}\right)^k t}\phi_k(t)\nonumber  \\
    &\le e^{-u}e
    \label{overlap-eq45}
\end{align}
which we get by choosing $t = (\ln{2}/ d)^k$. Note that P is the probability associated with the Poisson branching process. We will pick $u$ to make this small, but first we need to bound $Z_1 + Z_2 + \ldots + Z_k = B_k$ since this is what we compare with the graph neighborhood. Note that
\begin{align}
\mathrm{P} \Big[B_k \ge \lambda\Big] \le \mathrm{P}
\left[Z_1 \ge {\lambda\over k}\right]
+ \mathrm{P} \left[Z_2 \ge {\lambda \over k}\right] + \ldots + \mathrm{P} \left[Z_k \ge {\lambda \over k}\right].
\label{overlap-eq46}
\end{align}
Choose $\lambda = d^{sk} \left(d / \ln{2}\right)^k$ and $u = {d^{sk} \over k}$ with $s$ to be determined later. Then (for $k$ large enough, using $d>1$)
\begin{align}
\mathrm{P}\left[B_k \ge d^{sk} \left(d /\ln{2}\right)^k
          \right] \le k \, e^{-{d^{sk}\over k}} \, e
\label{overlap-eq47}
\end{align}
(since the moment generating function for $Z_k$ is the biggest) which we can write as
\begin{align}
\mathrm{P} \left[B_k \ge d^{sk} \left(d /\ln{2}\right)^k
          \right] \le  e^{-{d^{sk/2}}} 
\end{align}
for $k$ large enough.

This bound applies to Poisson branching. To make contact with our graph neighborhoods we first compare the branching process to Erdos-Renyi graphs and then to our random graphs with a fixed number of edges. In the Erdos-Renyi graph, each vertex has Binomial$(n-1, {d\over n-1})$ neighbors which as $n$ goes to infinity is Poisson($d$).  For finite $n$  the moment generating function of this Binomial is less than the moment generating function of the Poisson, so the $k$-neighborhood of a vertex in the random Erdos-Renyi graph satisfies the same bound as the branching process.
Let $f_y$ be the probability that (say) vertex $i$ in the Erdos-Renyi graph has its $k$-neighborhood at least as big as $d^{sk} ({d/\ln{2}})^k$, {\it conditioned} on the graph having exactly $y$ edges. (The number of edges is a Binomial $(\binom{n}{2},{d\over n-1})$ random variable.)
Then 
\begin{align}
\sum\limits^{\binom{n}{2}}_{y=0} \ \mathrm{P} (y \ \text{edges}) \, f_y \le \, 
e^{-d^{sk/2}}.
\label{overlap-eq48}
\end{align}
Starting the sum at $m={nd\over 2}$ we have
\begin{align} &\sum\limits^{\binom{n}{2}}_{y=m} \mathrm{P} (y \ \text{edges}) \, f_y \le e^{-d^{sk/2}} 
\end{align}
and since $f_y$ is an increasing function of $y$ we have
\begin{align}
	           &\sum\limits^{\binom{n}{2}}_{y=m} \mathrm{P} (y \ \text{edges}) \, f_m \le e^{-d^{sk/2}}.
\label{overlap49}
\end{align}
Now $m = {nd \over 2}$ is the expected number of edges so this is $({1\over 2} + o (1)) f_m$ and we conclude that 
\begin{align}
f_m \le (2 + o (1)) e^{-d^{sk/2}} \le e^{-d^{sk/3}}
\end{align} for $k$ large enough.
So using the indirect connection between the branching process and our graphs, we have
\begin{align}
\mathrm{Prob} \left[\mathrm{B} (i, k) \ge d^{sk} \left(d/\ln{2}\right)^k\right] \le e^{-d^{sk/3}}
\end{align}
where Prob is with respect to the graph distribution with a fixed number of edges $m$. 
Now take $k = {w \log n \over \log (d/ \ln{2})}$ and recall that by assumption $2p\le k$ and we have
\begin{align}
\mathrm{Prob}  \left[\mathrm{B} (i, 2p)\ge \left(d^s d / \ln{2} \right)^{w \log n \over \log (d/ \ln{2})}\right]\le
\mathrm{Prob}  \left[\mathrm{B} (i, k)\ge \left(d^s d / \ln{2} \right)^{w \log n \over \log (d/ \ln{2})}\right] \le e^{-d^{  s w \log n \over 3 \log (d/\ln{2})}}
. 
\end{align}
We can take $\log = \log_d$ in the above, and writing $\log_d (1/\ln{2}) = L$ we get
\begin{align}
\mathrm{Prob}  \left[\mathrm{B}(i, 2p)\ge n^{(1 + s + L) w \over 1 + L}\right] \le e^{-n^{s w  \over 3(1 + L)}}.
\end{align}
We can pick $s > 0$ to make $A={(1 + s + L) w \over 1 + L} < 1$ because $w < 1$. By the union bound, $\mathrm{Prob}[\max_i\mathrm{B}(i,2p)\ge n^A]$ can be at most $n$ times as large, so choosing $a<{sw\over 3(1+L)}$ yields the first half of the theorem. The other half, for $B (i, p)$, has $w\over 2$ in the exponent.  End of Proof.
\subsection{Concentration}

We now restate and prove the Concentration Theorem. This result concerns the concentration of the Hamming weight in the output strings of the shallow depth QAOA+.  The first part applies to the QAOA+ acting on a fixed random graph.  The second part is a statement about concentration on random graphs.

\vspace{10pt}
\noindent\textbf{Concentration Theorem}
\vspace{5pt}

\textit{1. Let $G$ be a graph with  $n$ vertices and ${d n\over2}$ edges. Suppose $p$ is chosen such that  $\max_{i }|\mathrm{B}_{G_n}(i,2p)|\le n^A$ for some $0 < A < 1 $.  Let $\sigma$ be the output of the QAOA+. 
Then there exists $\gamma_1>0$ such that for all $\delta>0$}
\begin{align}
\label{c1b}
\pr_{G}\Big[~\big ||\sigma|-\mathbb{E}_{G}|\sigma|\big|\ge \delta n~\Big]\le e^{-\delta n^{\gamma_1}}
\end{align}
\textit{ for $n$ large enough. Here the graph is fixed and the randomness comes from the QAOA+.}

\textit{2.  Now let $G$ be a random $\G(n,{dn\over2})$ graph. Suppose} 
\begin{align}
\label{c2b}
&\mathrm{Prob}\Big[\max_i\mathrm{B}(i,2p) \ge n^{A}\Big] \le e^{-n^a} \  \\and~~~
& \mathrm{Prob}\Big[\max_i\mathrm{B}(i,p) \ge n^{A/2}\Big] \le e^{-n^{a/2}}.
\end{align}
\textit{Then there exists $\gamma_2>0$ such that for any $0<\delta<1$},
\begin{align}
\label{c3b}
\mathrm{Prob}\Big[~\big| |\sigma|-\mathrm{Ex}|\sigma|\big|\ge \delta n~\Big]\le e^{- {\delta^2} n^{\gamma_2}},
\end{align}
\textit{for $n$ large enough. Here $\mathrm{Prob}$ and $\mathrm{Ex}$ refer to Probability and Expectation coming from the distribution of random graphs combined with the randomness coming from the QAOA+}.

\vspace{10pt}

Consider the QAOA+ running on a graph $G$ and outputting an independent set $\sigma$. The Hamming weight of $\sigma$ is a sum of $n$ random variables $b_1 + b_2 + \cdots + b_n $, where each $b_i \in \{0,1\}$ is independent of the collection $\{b_j\}$ for $j$ not within a distance $2p$ of $i$ in $G$.
With $p$ small enough so that the $2p$-neighborhoods  are small compared to $n$, this is enough to get a sub-exponential bound on the concentration of $| \sigma |$ for a fixed graph $G$. We use the standard technique of moment generating functions (as in the Chernoff bound) to prove \eqref{c1b}. 

Suppose  $\max\limits_i \mathrm{B}_G (i, 2p) \leq n^A$, with $A < 1$. Let
\begin{align}\label{eq:chernoff}
Y_i = b_i - \mathbb{E}_G\, (b_i).
\end{align}
We will first bound $\mathbb{E}_G\, [\sum Y_i]^t$ and use that to bound $\mathbb{E}_G[ e^{\theta \sum Y_{i}}]$, from which \eqref{c1b} will follow.
We have
\begin{align}\label{eq:firstbound}
\mathbb{E}_G\, \big[\sum Y_i\big]^t =\sum_{1\leq i_1, \cdots ,{i_t} \leq n} \, \mathbb{E}_G \big[Y_{i_{1}} \cdots Y_{i_{t}}\big].
\end{align}
Because each $\mathbb{E}_G[Y_i] = 0$, any term in the sum is 0 by independence unless for each $i_k$, there is an $i_\ell ~ (\ell \ne k)$ in $ \mathrm{B}_G(i_k,2p)$.
The number of nonzero terms is at most $t ! (n n^A)^{t/2}$ - see note at end of this section. Since $ | Y_i | \leq 1$, we have
\begin{align}\label{eq:nonzero}
\mathbb{E}_G\, \big[\sum \, Y_{i} \big]^t \ \leq \ t ! (n n^A)^{t/2}\ .
\end{align}
Multiply both sides by $\theta^t$, divide by $t!$ and sum on $t$ to get
\begin{align}\label{thetasupt}
\mathbb{E}_G\big[\, e^{\theta \sum Y_i}\big] \leq \sum^\infty_{t=0} \, \theta^t (n n^A)^{t/2} = \frac{1}{1-\theta n^{\frac{1 +A}{2}}} .
\end{align}
Now we can prove \eqref{c1b}. By the Markov inequality, for any $\theta$, 
\begin{align}
\pr_{G} \Big[\sum  Y_i \geq \delta n\Big] \leq e^{- \theta \delta n} \mathbb{E}_G\Big[ e^{\theta \sum Y_i}\bigg]
\end{align}
which combines with the previous result to give
\begin{align}
\pr_{G} \Big[\sum  Y_i \geq \delta n\Big]
\leq \frac{e^{- \theta \delta n}}{1-\theta n^{\frac{1 +A}{2}}}.\nonumber \hspace{3.5em}
\end{align}
Choose $0 < \gamma^\prime < \frac{1 - A}{2}\ $ and $\ \theta = n^{\gamma^\prime - 1}$, so
\begin{align}
\pr_{G} \Big[\sum \, Y_i \geq \delta n\Big] \leq  \frac{e^{-\delta n^{\gamma^\prime}}}{1 - n^{\left(\gamma^\prime - \frac{1}{2} + \frac{A}{2}\right)}}.
\end{align}
The bound on $\pr_{G} \big[\sum \, Y_i \leq - \delta n\big]$ is the same (use  $\ \theta = - n^{\gamma^\prime - 1}$). Replace $\gamma^\prime$ by any $\gamma_1 < \gamma^\prime$ and we have
\begin{align}
\pr_G\left[\left|\sum Y_i\right|\ge \delta n\right]\le e^{-\delta n^{\gamma_1}}
\end{align}
for large $n$. This is \eqref{c1b}.

The second concentration bound \eqref{c2b} follows from the fact that $\mathbb{E}_G | \sigma |$ is very unlikely to vary much as $G$ varies as we now show. The random graph $G$ consists of $m$ edges $e_1, e_2, \cdots e_m$ chosen independently uniformly over the $n\choose 2$ possibilities. 
(We ignore the $O(1)$ collisions.) Let
\begin{align}
f(\mathbf{e}) = f (e_1, e_2, \cdots e_m) = \mathbb{E}_G | \sigma |.
\end{align}
Here the graph is fixed and the expectation is with respect to  the QAOA+. As long as the $p$-neighborhoods of the vertices in $G$ are small, a change in one edge of $G$ makes only a small change in $f(e_1, e_2, \cdots e_m)$ by the locality of the QAOA+. It is natural to use the Azuma inequality which bounds concentration probability in terms of local changes. But the neighborhoods are only small with high probability so we cannot use the Azuma inequality directly on the function $f$. Instead we adjust $f$ using a version of Kirzbraun's Theorem on extending Lipschitz functions.

To set the stage we state the Azuma inequality. Consider a real valued function $\phi(\bf{e})$ where $\bf{e}$ is a set of $m$ variables $e_1, e_2 \cdots e_m$ with the property that $\phi$ does not change much when one of the variables changes.  Let $\tilde{\textbf{e}}$ be equal to $\bf{e}$ except in one of the $m$ variables with the Lipschitz condition on $\phi$ being
\begin{align} |\phi(\textbf{e})-\phi(\tilde{\textbf{e}})| \le R
\end{align}
for some fixed $R$. Now we take the $\{e_i\}$ to be independent random variables.  The Azuma inequality states that
\begin{align}\label{Azuma} \mathrm{Prob}\Big[ \big|~ \phi - \mathrm{Ex}[\phi]~\big| \ge t\Big]\le2 \mathrm{exp}\Big(\frac{-t^2}{2 m R^2}\Big).
\end{align}

Return to our function $f(\bf{e})$ where the $e_i$ are edges in a graph. We will use (78).  Let $\mathrm{K}_n$ be the set of graphs with $n$ vertices and ${d n\over2}$ edges that have small neighborhoods
\begin{align} \mathrm{K}_n=\big\{G:\max_i|B_G(i,p)|\le n^{A/2}\big \}
\end{align}
so
\begin{align}
\label{probKn}
\mathrm{Prob}\big[K_n\big]\ge 
1-e^{-n^{a/2}}
\end{align}
for $n$ large enough.
If $\bf{e}$ and $\tilde{\bf{e}}$ differ by one edge and both are in $\mathrm{K}_n$ the we have
\begin{align}\label{fminusf}
\big|f(\textbf{e})-f(\tilde{\textbf{e}})\big|\le 4n^{A/2}
\end{align}
since the QAOA+ outputs can only differ on bits in the 4 neighborhoods of the vertices at the ends of the 2 swapped edges. For graphs not necessarily in $\mathrm{K}_n$ we need to modify $f$. Let $\rho(\bf{e},\tilde{\bf{e}})$ be the number of edge changes needed to turn $\bf{e}$ into
$\tilde{\bf{e}}$. By \eqref{fminusf} if $\textbf{e}$ and $\tilde{\textbf{e}}$ are in $\mathrm{K}_n$ then
\begin{align}\label{fminusfagain}
\big|f(\textbf{e})-f(\tilde{\textbf{e}})\big|\le 4n^{A/2}\rho(\bf{e},{\tilde{\bf{e}}}).
\end{align}
Now define for any $\textbf{e}$,
\begin{align}
g(\textbf{e})=
\min_{\textbf{e}^\prime\in\mathrm{K}_n}
\big[f(\textbf{e}^\prime)+4n^{A/2} \rho(\textbf{e},\textbf{e}^\prime)\big].
\end{align}
For $\textbf{e}\in\mathrm{K}_n$ we see that $g(\textbf{e})=f(\textbf{e})$ by \eqref{fminusfagain}. For any $\textbf{e}$ and $\tilde{\textbf{e}}$, we have
\begin{align}
\label{gminusg}
\big|g(\textbf{e})-g(\tilde{\textbf{e}})\big|\le 4 n^{A/2}\rho(\bf{e},{\tilde{\bf{e}}}). 
\end{align}
To see this, note that there is an $\textbf{e}^\prime\in\mathrm{K}_n$ with
\begin{align}
\begin{split}
g(\textbf{e})=f(\textbf{e}^\prime)+4n^{A/2}\rho(\textbf{e},\textbf{e}^\prime)
\\
g(\tilde{\textbf{e}})\le f(\textbf{e}^\prime)+4 n^{A/2}\rho(\tilde{\textbf{e}},\textbf{e}^\prime).
\end{split}
\end{align}
Subtracting these two and using the triangle inequality yields
\begin{align}
g(\tilde{\textbf{e}})-g(\textbf{e})\le 4n^{A/2}\rho(\textbf{e},\tilde{\textbf{e}}).
\end{align}
Repeat with $\textbf{e}$ and $\tilde{\textbf{e}}$ interchanged and \eqref{gminusg} follows. We are going to apply the Azuma theorem to $g$ so note that if $\textbf{e}$ differs from $\tilde{\textbf{e}}$ in just one edge then $\rho(\textbf{e},\tilde{\textbf{e}})$ is $1$ in \eqref{gminusg}. Then for any $\delta$ the inequality \eqref{Azuma} says
\begin{align}
\label{Ag}
\mathrm{Prob}\Big[\big|g-\mathrm{Ex}[g]\big|\ge\delta n/2\Big]\le 2 \mathrm{exp}\Big(\frac{-(\delta n/2)^2}{2(d n/2)(4n^{A/2})^2}\Big)
=2\mathrm{exp}\Big(-\frac{\delta^2 n^{1-A}}{64 d}\Big).
\end{align}

We are going to use this to bound $\mathrm{Prob}\big[|f-\mathrm{Ex}[f]|\ge \delta n \big]$ which appears in \eqref{c3b} since again $f=\mathbb{E}_G[|\sigma|]$ and $f$ agrees with $g$ except outside of $\mathrm{K}_n$. Outside of $\mathrm{K}_n$ we have the crude bound that $|f-g|\le (n + 4 n^{A/2} d n /2)$ so by \eqref{probKn}
\begin{align} \big|\mathrm{Ex}[f]-\mathrm{Ex}[g]\big|\le\big(n + 4 n^{A/2} d n /2 \big) e^{-n^{a/2}}\le \delta n / 2,
\end{align}
for any $\delta$ if $n$ is big enough. Now
\begin{align}
\mathrm{Prob}\Big[ \big|f-\mathrm{Ex}[f]\big|\ge \delta n\Big]\le\mathrm{Prob}\Big[\big|f-\mathrm{Ex}[g]\big|\ge \delta n / 2\Big].
\end{align}
Inside $\mathrm{K}_n$ we have $f=g$ while the probability of being outside of $\mathrm{K}_n$ is bounded by \eqref{probKn} so
\begin{align}
\mathrm{Prob}\Big[ \big|f-\mathrm{Ex}[f]\big|\ge \delta n \Big]\le\mathrm{Prob}\Big[ \big|g-\mathrm{Ex}[g]\big|\ge \delta n / 2\Big]+e^{-n^{a/2}},
\end{align}
which by \eqref{Ag} is
\begin{align}
\mathrm{Prob}\Big[
\big|f-\mathrm{Ex}[f]\big|\ge \delta n \Big]\le2\mathrm{exp}\Big(-\frac{\delta^2 n^{1-A}}{64 d}\Big)+e^{-n^{a/2}}.
\end{align}
We can rewrite this (take $\delta<1$)  as
\begin{align}
\mathrm{Prob}\Big[\big|\mathbb{E}_G|\sigma|-\mathrm{Ex}|\sigma|\big|\ge \delta n\Big]\le e^{-{\delta^2}n^{\tilde a}}
\end{align}
with $\tilde a$ less than both $1-A$ and $a/2$, for $n$ large. The first half of the concentration theorem implies
\begin{align}
\mathrm{Prob}\Big[\big||\sigma|-\mathbb{E}_G|\sigma|\big|\ge \delta n\Big]  \le e^{-\delta n^{\gamma_1}} + e^{-n^a}, 
\end{align}
where the $e^{-n^a}$ accounts for graphs with large $2p$-neighborhoods.

Now the triangle inequality says
\begin{align}
\mathrm{Prob}\Big[\big||\sigma|-\mathrm{Ex}|\sigma|\big|\ge 2 \delta n \Big] \le e^{-{\delta^2}n^{\tilde a}} +  e^{-\delta n^{\gamma_1}} +e^{-n^a}.
\end{align}
By taking $\gamma_2$ smaller than $\tilde a,\gamma_1$ and $a$ we obtain
\begin{align}
\mathrm{Prob}\Big[\big||\sigma|-\mathrm{Ex}|\sigma|\big|\ge 2 \delta n \Big] \le e^{-{\delta^2} n^{\gamma_2}},
\end{align}
for all $n$ large.  This is \eqref{c3b}, as desired, if we go back to the beginning and let $\delta$ go to $\delta/2$.

\subsection{Counting the number of non-zero terms}

To bound the number of non-zero terms in 
\begin{align}
\label{EG}
\sum\limits_{0 \le i_1, \ldots, i_t \le n}\ \mathbb{E}_G\left(Y_{i_1}\cdots Y_{i_t}\right)
\end{align}
first note that since each $\mathbb{E}_G[Y_i] = 0$, any term in the sum is 0 by independence unless for each $i_k$, there is an $i_\ell ~ (\ell \ne k)$ in $ \mathrm{B}_G(i_k,2p)$. We are going to graphically represent the non-zero contributions by considering graphs with vertices 
$1, 2, \ldots t$, possibly disconnected. A graph will be called valid if no vertex is isolated. We say  that a sequence $i_1, i_2, \ldots i_t$ satisfies a graph $\Gamma$ on $\{1, 2, \ldots t\}$ if  $i_\ell$ is in the $2p$-neighborhood of $i_k$ (and, equivalently, vice-versa) whenever $(\ell, k)$ is an edge of $\Gamma$. We know that every sequence $i_1, i_2, \ldots i_t$ with $\mathbb{E}_G\left(Y_{i_1}\cdots Y_{i_t}\right) \ne 0$ satisfies at least one valid graph, and in particular it satisfies some minimal valid graph -- minimal means no subgraph with $t$ vertices but fewer edges is valid. So we need to bound the number of minimal valid graphs as a function of $t$. Call this number $V_t$.
A minimal valid graph has this form: each of its components is a star, that is, a tree with a central vertex connected to all the other vertices in the component.
Suppose the largest component has size $r$. If $r=2$, then the minimal valid graphs are matchings ($t$ must be even) of which there are $(t-1)(t-3) \ldots 1$ for $t\ge 2$. 
For $r > 2$, there $t \binom{t-1}{t-r}$ stars of size $r$, and at most $
t \binom{t-1}{t-r}\ V_{t -r}$
minimal valid graphs which may be overcounting because we've singled out one component of size $r$. So for $t \ge 2$
\begin{align}
V_t  \le &\ (t-1)(t-3) \cdots 1 + \sum\limits^{t-2}_{r=3}\ t \binom{t-1}{t-r}\ V_{t -r} \nonumber\\
	=    &\  {t! \over t(t-2)\cdots 2} + \sum\limits^{t-2}_{r=3} \ {t(t-1)\cdots (t-r +1) \over (r-1) !}  V_{t-r}.
\end{align}
Now suppose inductively that $V_{t-r} \le (t-r)!$ for $r \ge 3$. (Note that $V_2 =1$ and $V_3 = 3$.) Then for $t \ge 4$,
\begin{align}
V_t  \le &\ {t! \over t(t-2)} + \sum\limits^{t-2}_{r=3}\ {t(t-1)\cdots (t-r +1) \over(r-1) !} (t-r)! \nonumber \\
	     \le &\ {t! \over 8} + \sum\limits^{\infty}_{r=3}\ {t! \over (r-1)!}\\ = &\ t! ({1\over 8} + e-2) < ~t!~.
\end{align}
So there are at most $t!$ minimal valid graphs with $t$ vertices. How many sequences $i_1, i_2, \ldots, i_t$, where $1\le i_\ell \le n$, can satisfy a given minimal valid graph? If it has $u$ components, then it must have $t-u$ edges. The central vertex $\ell$ in any component corresponds to $i_\ell$ which has $n$ possible values. If $(\ell, k)$ is an edge in the minimal valid graph, $i_k$ can have only $n^A$ values, since the sequence $i_1\ldots i_t$ is assumed to satisfy this minimal valid graph.
So there are at most $n^u (n^A)^{t-u}$ possible sequences and since $u\le t/2$, this is $\le n^{t/2} (n^A)^{t/2}$.
Multiplying by the bound on the number of minimal valid graphs, there are at most
\begin{align}
t! n^{t/2} (n^A)^{t/2}
\end{align}
nonzero terms in \eqref{EG}.

  \section {Discussion}
  
  No one knows if a quantum computer running a quantum algorithm will be able to outperform a classical computer on a combinatorial search problem.  One approach is to build a quantum computer, run a quantum algorithm and see what happens at the available number of qubits. Another approach is to look for a provable quantum speedup over the best known classical algorithms. To this end it is useful know if there are provable limitations to the power of any proposed quantum algorithm.  
  
  In this paper we look at the Quantum Approximate Optimization Algorithm applied to finding a large independent set in a random graph of fixed average degree $d$. The performance of the QAOA can only improve with depth $p$ but we show that for Maximum Independent Set on random graphs the algorithm will fail to pass a certain performance barrier if $2p$ is less than $w \log n /\log{ (d/\ln{2}) }$ for any $ w < 1$ with $d$ big enough.  (This ratio is independent of the base of the $\log$.) The quantum algorithm consists of $p$ unitaries that each respect the locality of the underlying graph.  With a fixed average degree of $d$ this means that each qubit typically has an influence sphere of roughly $d^p$ other qubits. For qubits further than $2p$ apart on the graph these influence spheres do not intersect and we can show that measurements of these qubits are uncorrelated. This is key to our showing that the algorithm has limited power.  However if $p$ is large enough that $d^p$ exceeds $n$ our arguments do not apply and we have no indication that the QAOA will fail.
  
  Our results are for random graphs and we do not have results for say a graph which is a $2$ dimensional square lattice.  In this case perhaps the border between failure and possible success is at $\sqrt{n}$.  Back to random graphs. Although our proof technique requires $d$ big, the intuition is that if $d^{2p} < n$ then most pairs of qubits will have independent measurement outcomes.  Consider the case when the degree is small so the influence spheres are small, the least favorable situation for the QAOA. For example with $d=3$ at one million qubits  our result sugggest  failure for $p$ less than $7$. The bipartite construction of \cite{BKKT} only shows failure at 2 million qubits with $d=3$  for $p$ less than $1$. Just beyond this the QAOA ``sees" the whole graph and we can not say with certainly what happens at a few million qubits in the shallow circuit depth regime with $p$ say in double digits.  
  
  \section*{Acknowledgement}
  We thank Aram Harrow for helpful discussion.

\bibliographystyle{unsrt}
\bibliography{bibliography}

\end{document}